\documentclass{cimento}

\usepackage{mathrsfs}
\usepackage{amsmath}
\usepackage{amssymb}
\usepackage{color}
\usepackage{graphicx}
\usepackage{color}
\usepackage{epsfig}

\title{Cosmic Rays, Gamma rays, Neutrinos and Gravitational Waves}

\author{Paolo Lipari \from{inst1}}

\instlist{\inst{inst1} INFN, Sezione di Roma ``Sapienza'', Italy}

\begin{document}

\maketitle

\begin{abstract}
This paper discusses the relation between the study of the
fluxes of cosmic rays, gamma rays and neutrinos, and the connection
of these observations with the newly born field of gravitational wave astronomy.
\end{abstract}

\section{Multi--messenger Astrophysics} 
The discovery of cosmic rays at the beginning of the
20th century was the first glimpse of  what is now known as
the ``High Energy Universe'', the ensemble
of the astrophysical objects, environments and mechanisms 
that generate or store very high energy particles. In recent years
our understanding of these phenomena has made great progress
thanks to studies performed using three ``messengers'':
cosmic rays, gamma rays and neutrinos.
Cosmic rays (CR) are relativistic, electrically charged particles of
different types: protons, nuclei, electrons,
with smaller but very important contributions of antiprotons and positrons.
The studies of these three messengers are intimately connected,
and should be considered as the three faces
of a single scientific field.

The relation between CR's , $\gamma$'s and $\nu$'s  is 
simple: the dominant source of high energy $\gamma$'s and $\nu$'s 
is emission from CR particles. 
Gamma rays can be generated by relativistic hadrons
(protons and nuclei) or charged leptons (electrons and positrons).
In the first case the emission mechanisms are bremsstrahlung
and Compton scattering (where the targets are the soft photons
that form the radiation fields in the medium where
the $e^\mp$ are propagating).
In the second case the photons are generated in the
decay of neutral pions ($\pi^\circ \to \gamma \gamma$) and other unstable mesons 
created in the inelastic interactions of relativistic protons and nuclei,
with a target that can be a gas of ordinary matter, or a radiation field.
The hadronic mechanism is also a neutrino source, 
because the final state of hadronic interactions also contains
particles that decay into $\nu$'s. The dominant channel
is the chain decay of charged pions
[$\pi^+ \to \mu^+ \nu_\mu \to (e^+ \, \nu_e \, \overline{\nu}_\mu) \; \nu_\mu$
and charge conjugate modes]. 
The rates of production
of the three pion states are approximately equal, and therefore 
the $\nu$ and $\gamma$ emissions are of approximately the same size.

The only significant source of CR's
that exists  with certainty is the acceleration of electrically
charged particles in astrophysical objects
(or better ''events'', since in many cases the sources are transient).
The interactions of these accelerated primary
CR particles ($p$, $e^-$ and most nuclei) 
can then generate gamma rays and neutrinos, and also
other secondary CR particles (such as $e^+$, $\overline{p}$ and 
rare nuclei such as beryllium or boron).
Understanding the properties of the high energy astrophysical sources
is a fundamental goal of multi--messenger astrophysics.

The formation of electromagnetic fields sufficiently strong
and extended to be able
to accelerate particles to relativistic energies are often
associated to violent astrophysical events
(such as gravitational collapses or mergings of compact objects)
where large masses undergo large accelerations.
These events are therefore
also very powerful sources of gravitational waves (GW),
and are very important targets for the GW telescopes.
The conclusion is that the high energy universe can be studied
not only the three messengers already listed
(CR's, $\gamma$'s and $\nu$'s), but also with a
fourth messenger: gravitational waves,
that can give unique information about the formation of the sources.

It is possible, that acceleration is not the only source
of very high energy particles and that a
non ngligible fraction has a different origin.
If the dark matter (DM) is in the form of Weakly interacting massive particles (WIMP's),
the self annihilation or decay of these particles generate
fluxes of secondaries with an energy spectrum that extends to a maximum energy
$E_{\rm max} \simeq M$ ($M/2$ in the case of decay).
Theories based on Supersymmetry do predict the existence of a stable particle
with a mass, that (assuming the validity of the concept of ``naturalness'')
should not be much larger that the Weak mass scale
(implying a mass for the DM particle of order $M \sim 10^2$--10$^4$~ GeV).
Supersymmetric scenarios are currently the object
of a large and varied program of experimental studies.
Another, more speculative possibility is that
the universe contains super massive particles, with a mass of
order of the Grand Unification scale ($M \gtrsim 10^{24}$~eV),
that could perhaps also form a part or all of the the dark matter.
If these super--massive particles exist with a sufficiently
large density and are unstable,
they could generate observable fluxes of ultra high energy particles.

\begin{figure}
\includegraphics{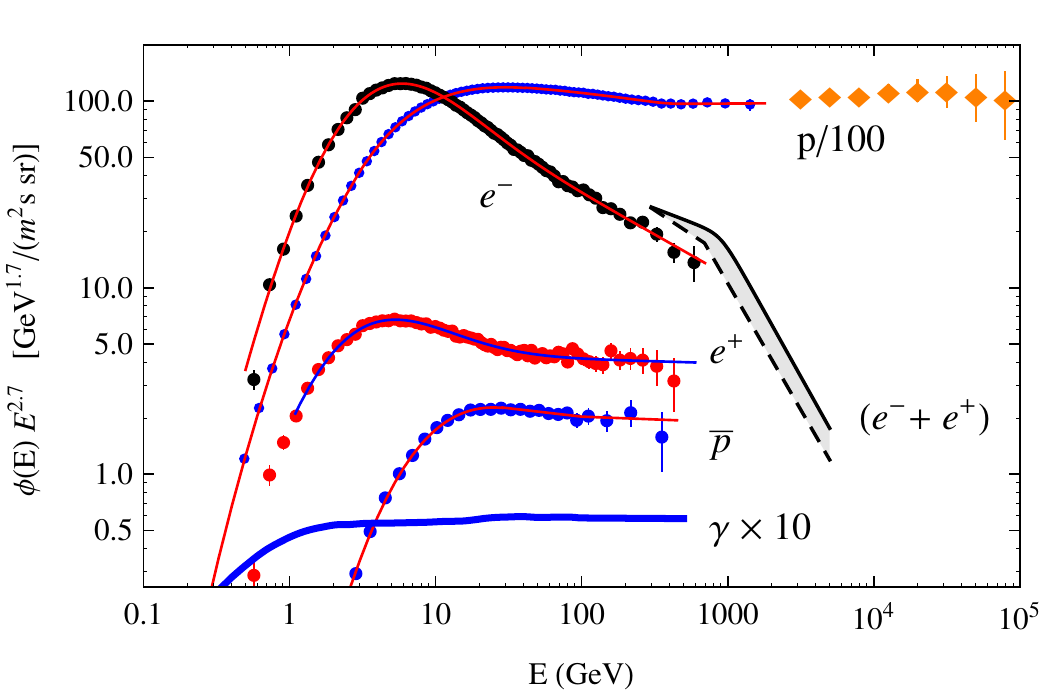} 
\caption{
\label{fig:low-cr}
\footnotesize
Energy spectra of different types of CR particles
($p$, $e^\mp$ and $\overline{p}$) (by AMS02 and CREAM).
The measurements of the ($e^+ + e^-$) flux have
been obtained by the ground--based Cherenkov telescopes (HESS, MAGIC and VERITAS).
The average diffuse Galactic gamma ray flux measured by the FERMI
telescope is also shown.
The $p$ and $\gamma$ spectra have been rescaled.
For references see \cite{Lipari:2016vqk}. 
}
\end{figure}

\section{Cosmic Rays}
It is important to discuss the present and future
CR studies in the general context of multi--messenger astrophysics,
and in particular together with the informations obtained
from the observations of $\gamma$ and $\nu$ telescopes.
Gamma rays and neutrinos propagate along straight lines, and
therefore the angular distribution of their flux maps the space
distribution of the emission.
This allows to perform detailed studies of individual point like
or quasi--point like sources, and to follow their time evolution
(when it is sufficiently fast).
In the last decades, gamma ray observations in the GeV and TeV
energy ranges have been living through a real ``golden age'',
yielding important results that have deeply transformed
our understanding of the high energy universe.
Telescopes  in space and at groud level have identified
a large number of sources (over 3000 in the GeV range
and approximately 200 in the TeV range) that
belong to several distinct classes of astrophysical objects
(Supernova remnants, Gamma Ray Bursts, Pulsar Wind Nebulae,
Microquasars, Active Galactic Nuclei, $\ldots$). The
study of the structure and properties of these objects is
currently a very rich field, with many open problems.

In contrast, the trajectories of (electrically charged) CR's
are bent by magnetic fields, and this results
in fluxes that are approximately isotropic and carry 
very little information about the space and time distributions of the sources.
This is clearly a very significant limitation, 
however, the energy spectra of the CR particles
encode very valuable information about the CR sources that
is important for the understanding of the high energy universe
and that is difficult to infer from $\gamma$ and $\nu$ observations, as
they are determined by the space and time  averaged spectra
of the released by the sources.
The CR fluxes are also shaped by the
properties of propagation of relativistic, electrically charged particles
in Galactic and extragalactic space, and therefore also give
unique information about the astrophysical magnetic fields.

A very compact summary of the main properties of the CR energy spectra
is contained in Figs.\ref{fig:low-cr} (for lower energy)
and~\ref{fig:high-cr} (for higher energy).
A first, obvious, but very important point is that the CR spectra extend
up to $E \simeq 10^{20}$~eV, an energy six order of magnitude larger that the
highest observations of existing Cherenkov telescopes. The fact 
that the CR sources are capable of generating particles of such
high energy is an essential constraint for the modeling of the sources.

\subsection{Galactic cosmic rays}
The flux of Galactic CR of type $j$ observable
at point $\vec{x}$  in the Galaxy can be written as the product:
\begin{equation}
\phi_j (E, \vec{x}) \simeq \left \langle Q_j (E) \right \rangle ~ P_j(E, \vec{x})
\label{eq:cr-decomposition}
\end{equation}
where $\langle Q_j (E) \rangle$ is the space integrated
and time averaged rate of release of CR particles of type $j$ and energy $E$ in the entire Galaxy,
and $P_j (E, \vec{x})$ is a ``propagation function''
(with dimension time divided by volume) that encodes the residence time and confinement
volume of CR of type $j$ and energy $E$, and
depends on the properties of CR propagation in the Galaxy\footnote{
In principle, it is possible that
the observable CR flux receive significant contributions
from one (or a few) near, young sources. In this case
the  (time averaged) stationary solution for Eq.~(\ref{eq:cr-decomposition}) is not
a viable model.}.
This  equation illustrates the fundamental difficulty in the interpretation of the 
CR flux measurement: the disentangling of the effects of the sources and
of propagation in the formation of the observed spectra.

The determination of the (time  averaged) source spectra
for CR of  different types:
$\langle Q_{p} (E) \rangle$, $\langle Q_{e^-} (E) \rangle$, $\ldots$ 
released in interstellar space  by the Galactic sources
is  obvious a problem of great importance.
The most commonly accepted theory is that most of the Galactic CR's are accelerated
in young SNR's, but there are also alternative hypothesis.
It has  been for example proposed that the dominant source of CR  are GRB's, 
and an alternative possibility is that the main  CR source  is  a single object,
the supermassive black hole at the  Galactic Center, during
periods of enhanced activity.

Considering the ``standard model'' of CR acceleration in SNR's,
at the present time the Milky Way contains a finite number
of sources, each emitting spectra of $\gamma$'s and $\nu$'s
that are  determined by the populations of relativistic particles contained in
each  source and by the properties and  structure
of the object (such as gas density or  magnetic fields).
The $\gamma$ and $\nu$   emissions are  directly observable at the Earth, 
but is far from easy, and strongly model dependent,
to estimate from the ensemble of these  observations,
(that capture a single ``snapshot'' of the evolution for  each source)
the time integrated spectra of cosmic rays
[$N_{p}^{s, {\rm out}} (E) $, $N_{e^-}^{s, {\rm out}} (E)$, $\ldots$]
that each source $s$  will release in interstellar space during its activity.
In fact,  the observations show the existence
very large differences in the $\gamma$ emission of different SNR's
that very likely reflect not only the differences in the properties
of the sources and their environments, but also a strong time dependence of the emission.
The measurements of the CR fluxes, that are determined by the space and time averaged
spectra  released by the sources are therefore of  great importance
to  constrain the modeling of the sources. 
This  study however requires a sufficiently good understanding of the properties of CR 
propagation  in the Galaxy  to  estimate the Galactic  source  spectra from
the CR observations.

Some crucial problems emerge naturally.
The first one is the determination of  the exponent
of the source spectra for protons, nuclei and electrons.
In the energy range $E \sim 10$--10$^3$~GeV the proton flux
is reasonably well described by a power law with exponent $\alpha_p \simeq 2.7$--2.8,
while the electron spectrum is significantly softer with an
exponent $\alpha_e \simeq 3.1$--3.2 (see Fig.~\ref{fig:low-cr}).
If  the shape of the observable flux is modeled 
as the softening of a  (power law) source spectrum   because of propagation effects,
the spectral indices of protons and electrons  can be written as the sum:
$\alpha_p \simeq \alpha_p^0+ \delta_p$, and $\alpha_e \simeq \alpha_e^0 + \delta_3$,
where  $\alpha_{p,e}^{0}$ is the index of the source spectrum,
and $\delta_{p,e}$ describes the energy (or rigidity) dependence of the propagation effects.
A crucial task is the determination of
the values of the exponents $\delta_p$  and $\delta_e$,
to establish if the  difference in spectral shape between protons and electrons
is determined by the sources, or by propagation effects.

A second problem that has an answer with broad and profound implications is the origin of the
so called ``knee'' in the all particle flux at energy $E_{\rm knee} \simeq 3 \times 10^{15}$~eV
(see Fig.~\ref{fig:high-cr}). It is common to to interpret this spectral
structure as the manifestation of a  maximum acceleration energy in the CR sources 
that  has values confined to a very narrow range  
The alternative hypothesis is that this knee is imprinted by
propagation effects on a smooth source spectrum.
The implications for the modeling of the sources are  obviously  very important.
In the first case  the Galactic CR  sources must  ``Pevatrons'',
reaching a maximum energy proton  enegy of  order 3~PeV,
approximately equal in all objects.
In the second  case, the CR sources must be capable of
reaching  sigificantly higher energies, challenging theorists.


\subsection{Positron and antiproton spectra}
The study of the $e^+$ and $\overline{p}$ spectra
is probably the topic in CR  studies that has received most attention in recent years.
This strong interest is a consequence of the fact that 
the antiparticle spectra are important probes to study the 
existence of DM in the form of WIMPs. Also  the intriguing
possibility that $e^+$'s  are  created  and accelerated in astrophysical sources
has been considered by several authors.

In the absence of such ``exotic''  mechanisms, the main source
of CR antiparticles is  their creation in the inelastic interactions
of primary particles.  The crucial question  is then  to  determine if
the observed  $e^+$ and $\overline{p}$  fluxes   are (or are not) consistent
with the hypothesis that the standard mechanism of  production
is their only source.
It is  intriguing   \cite{Lipari:2016vqk}
that the observed   fluxes of $e^+$ and $\overline{p}$
have the spectral shapes and the relative  normalization
that are  consistent with the hypothesis  that:
(i) they are created by the standard mechanism, and
(ii) the effects of propagation are  equal for the  two antiparticles
[$P_{e^+}(E) \approx P_{\overline{p}}(E)$].
These  result  can be most easily seen in Fig.~\ref{fig:low-cr} noting
that the $e^+$ and $\overline{p}$ fluxes for $E \gtrsim 20$~GeV, are power laws
with  the same   spectral  index and a ratio $e^+/\overline{p} \approx 2$ that is
equal to the ratio of the source spectra.
The simple, natural interpretation of this  result
is that  $e^+$ and $\overline{p}$ are both of secondary origin,
and that the propagation effects are small in  size and approximately equal  for
both antiparticle types.
Since  the rate of energy loss  rate of relativisitic $e^\pm$ in much larger
than the loss rate of $p$ and $\overline{p}$ (because
of the contributions of synchrotron and Compton losses),
this  implies that the residence time of $e^\pm$ is  sufficiently
short so that the total energy loss remains negligibly small.
It then also  follows that the different spectral shape of $p$ and $e^-$
is  generated by the sources and not determined by propagation effects.
The alternative possibility  requires  the  introduction of a new hard source of positrons.
The clarification of this  problem has broad and profound consequences.

\subsection{Galactic/extragalactic cosmic rays}
It is obviously of great importance to separate the populations of
CR's generated by sources in our Galaxy and by extragalactic sources.
It is firmly established that most of the lower energy particles 
are of Galactic origin, and it is also very likely that the highest energy
CR's are extragalactic. It remains controversial
what is the ``transition energy'' $E^*$ where the Galactic and extragalactic
components are approximately equal.

\begin{figure}
\includegraphics{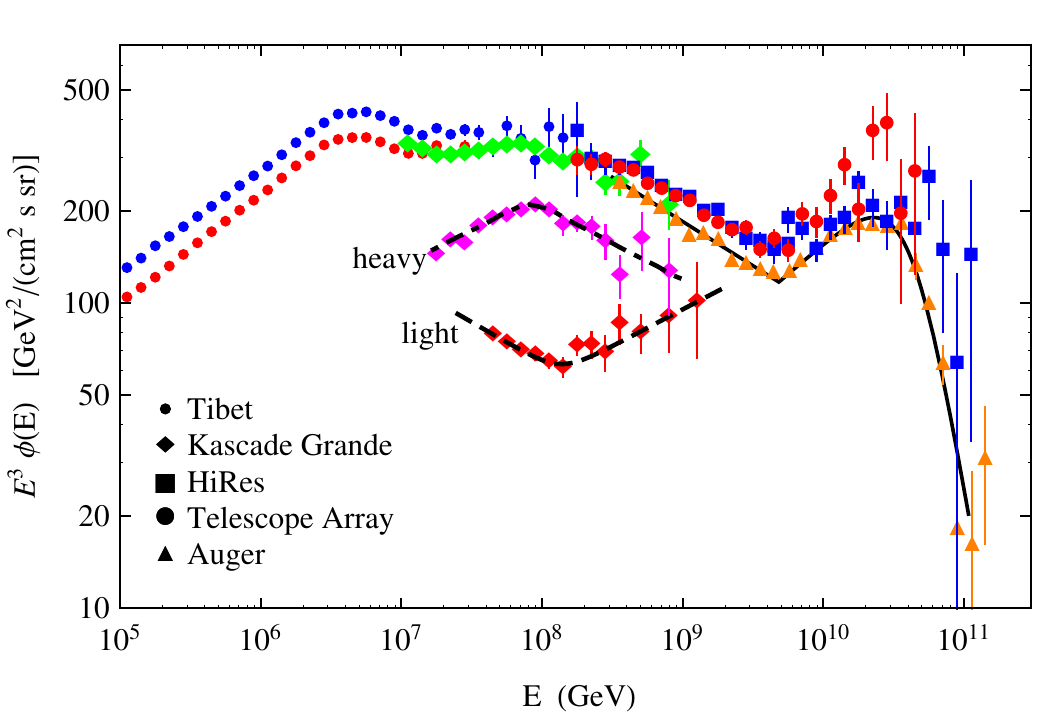} 
\caption{
 \label{fig:high-cr}
\footnotesize
 Measurements of the all particle CR flux 
(shown in the form $E^{3} \; \phi(E)$ versus $E$)
 For the Kascade--Grande data also
 measurements of the fluxes of events selected as ``light'' ($p$ rich)
 and ``heavy'' (iron rich) are shown.
 For references see  \cite{Lipari:2015ppa}.
}
\end{figure}

The determination of  $E^*$ is a task of crucial importance, because it
provides an essential  constraint on the properties of the
Galactic and extragalactic CR sources.
It is natural to expect that at the transition energy the CR energy spectrum
should correspond to an hardening of the CR spectrum.
Inspecting the all particle spectrum in Fig.~\ref{fig:high-cr} one can see that
there is only one significant hardening feature in the all particle CR flux,
the so called ``ankle'' at $E_{\rm ankle} \simeq 4 \times 10^{18}$~eV.
The identification of $E_{\rm ankle}$  with $E^*$ implies that some
Galactic sources must be capable to accelerate CR  up to very high energy.
Several  theories predict  however that $E^*$ is at lower energy, anmd  corresponds
to a softening spectral feature.
This requires some ``fine tuning'' of the shapes of the two components
or the existence of some, yet not understood physical mechanism, to  have
a sufficiently smooth transition.
In the popular ``dip model'' introduced by Berezinsky and collaborators, the ankle
is interpreted as an absorption feature (due  to the effects of  pair production interactions
$p \gamma \to p e^+e^-$ acting on a proton dominanted extragalactic flux), 
and the extragalactic compoment  becomes dominant at 
$E \simeq 10^{17}$~eV, the energy of the so called ``second knee''.
<
\subsection{CR Anisotropies}
The study of the structure and rigidity dependence
of the large scale CR anisotropies
is a very important tool to understand the properties of
CR confinement in the Galaxy.
However the deviations from isotropy of the CR angular distributions are
small and difficult to measure. The  anisotropies  remain below (or close) to
the level of $10^{-3}$ at lower  energy, and reach 
the level of $\sim 1\%$ for $E \simeq 10^{18}$~eV and 
$\sim 4\%$ at the highest energies ($ E \gtrsim 8 \times 10^{18}$~eV).
These effects remain unfortunately poorly understood.

An attractive possibility is that 
at very high rigidity ($E/Z \gtrsim 10^{19}$--10$^{20}$~eV)
the CR magnetic deviations,
even for propagation across extragalactic distances,
are sufficiently small  to allow the identification of the sources,
opening the window of ``proton astronomy''.
Some intriguing hints of correlations between the
directions of very high energy events 
and the positions of near Active Galactic Nuclei (AGN)
have been obtained by the Auger observatory,
but the results remain inconclusive.
Another interesting effect is possible signal (with 1.4\% of probability) of an
excess of events in a cone of 15$^\circ$ around 
the direction of Centaurus A, that a distance of 4~Mpc is AGN closest
to the Earth.
One can conclude that very large exposures at very high energy
are necessary for the unambiguous imaging of CR sources.

%


\begin{thebibliography}{100}
\bibitem{disclaimer} A more complete list of references can be found in the following two papers.
\end{thebibliography}
\end{document}